\documentclass[]{elsart}

\usepackage{graphicx}

\usepackage{amssymb}
\usepackage{latexsym}
\usepackage{graphs}

\begin{document}
\journal{Physica A}
\begin{frontmatter}



\title{Thesaurus as a complex network}


\author{Adriano de Jesus Holanda}, 
\author{Ivan Torres Pisa},
\author{Osame Kinouchi},
\author{Alexandre Souto Martinez\corauthref{cor1}} 
\corauth[cor1]{}
\ead{martinez@dfm.ffclrp.usp.br}
\ead[url]{http://www.fisicamedica.com.br/martinez/}
 \and \author{Evandro Eduardo  Seron Ruiz}

\address{Faculdade de Filosofia, Ci\^encias e Letras de Ribeir\~ao Preto (FFCLRP) \\
         Universidade de S\~ao Paulo (USP) \\
         Av.~Bandeirantes, 3900 \\
         14040-901  Ribeir\~ao Preto, SP, Brazil.
}

\begin{abstract}
A thesaurus is one, out of many, possible representations of term (or word) connectivity.
The terms of a thesaurus are seen as the nodes and their relationship as the links of a directed graph. 
The directionality of the links retains all the thesaurus information and allows the measurement of several quantities.  
This has lead to a new term classification according to the characteristics of the nodes, for example, nodes with no links in, no links out, etc. 
Using an electronic available thesaurus we have obtained the incoming and outgoing link distributions. 
While the incoming link distribution follows a stretched exponential function,
the lower bound for the outgoing link distribution has the same envelope of the scientific paper citation  distribution proposed by Albuquerque and Tsallis~\cite{albuquerque:2000}.
However, a better fit is obtained by simpler function 
which is the solution of Ricatti's differential equation. 
We conjecture that this differential equation is the continuous limit
of a stochastic growth model of the thesaurus network. We also propose
a new manner to arrange a thesaurus using the ``inversion method''.
\end{abstract}

\end{frontmatter}

Words are the building blocks to construct sentences and to transmit information. 
During last decades much effort has been spent on the statistics of words.  
Concern has been centered in the similarities and differences among word  distributions which may be useful for application in automatic information retrieval and thesaurus construction.  

Zipf~\cite{zipf_book} has shown that word frequency obeys a power law if words are ranked from the most to the less frequent ones. 
Statistical linguistics, at its lowest level, can be exemplified by
the Zipf's exponent, which is very sensitive to the writer's
instruction degree but much less sensitive to language (culture characteristics). 
Beyond word level, word connectivity has been treated in several manners. 
These treatments include entropic measures~\cite{montemurro:2001} and the construction of other quantities, such as the distribution of documents over the frequency of words~\cite{volchenkov:2003}.
Another interesting way to treat data is the Latent Semantic Analysis (LSA)~\cite{landauer:1997} which deals with word covariance in a corpus. 
LSA is a principal component analysis (PCA) technique , i.e., the covariance matrix is diagonalized and from the most important eigenvalues (around 300) the eigenvectors are considered to span an Euclidean vector space. 
A curious application of LSA is the automatic grading of high school texts~\cite{kintsch:2002}. 
However, LSA has been criticized as a poor approach for predicting semantic neighborhood~\cite{steyvers_griffiths}.

Other studies have focused on a different approach. 
Words are tied to each other as links of a graph where the words are the nodes of it.
Exhaustive studies over thesaurus~\cite{sigman,motter:2002} indicate that words are related among themselves as a small-world and scale-free network~\cite{watts}. 
This means that words may be embedded in a low dimensional space but with a small fraction of long distance connections. 
The existence of the low dimensional space has been suggested by the deterministic ``tourist'' walks ~\cite{lima_prl2001,stanley_2001} on the graph, which is an independent sampling procedure~\cite{kinouchi_phys_a}.

%

A \emph{thesaurus} is a list of terms.  
A \emph{term} can be a word, a composed word or even an expression.  
The list of related terms to a main entry term (head-word) provides alternatives for these entries.
Following previous studies, we will consider terms as being ``words'' in a broad sense.

As in a previous work~\cite{motter:2002}, our study is based on unstructured thesaurus, the \emph{Moby Thesaurus II} which is the largest\footnote{The file has 24,271 KB.} and most comprehensive free thesaurus data source in English available~\cite{mthesaurus}.
It has 30,260 (main) \emph{entries}, also called \emph{root
  words}\footnote{A \emph{root word} should not be confused with a
  \emph{root node} which is defined as one that has no incoming link.}
or \emph{head-words}\footnote{Some curiosities are: 877 words which
  are not referred from other entries, 16 words are entry words but
  point only to non-root words.} and 73,046 words which are referred
from the entries but they are not entries. They are called \emph{non-root words}. 
These add up to 103,306 different words. 
Each root word points, in average, to 83 words\footnote{From which, in average, 54 are root words and 29 are non-root words.}. 

The thesaurus derived network is defined considering each term as a node. Connections are established from an entry to its related list of terms forming a \emph{directed graph}. 
Classification of terms can be accomplished looking at the links (arcs), for instance:  head words (root words) are words with at least one emerging link ($k_{out} > 0$) and  non-root words are words with no emerging links ($k_{out} = 0$). 
Apparently there is a giant strong component (percolative cluster of directed links) which connects a large fraction of words~\cite{newman:2001,dorogovtsev:2001}. 
We stress that the working thesaurus is a simple and unstructured \emph{related term} thesaurus and we point out the existence of other thesauruses such as WordNet~\cite{wordnet} and definition terms thesaurus (Roget's thesaurus) which may be modeled as a bipartite graph, but they are be considered here. 

If only \emph{co-linked terms} (mutually referred terms) are
considered, this structure forms a digraph and reduces to the previous
studied one, where its small-world and scale free  structure has been pointed out~\cite{motter:2002}.  
In this case, the number of connections $k$ of a node is called \emph{degree of a node}. The node degree statistics shows an exponential behavior for small values of $k$ and a power law behavior for large values of $k$~\cite{motter:2002}.

\begin{figure}[ht]
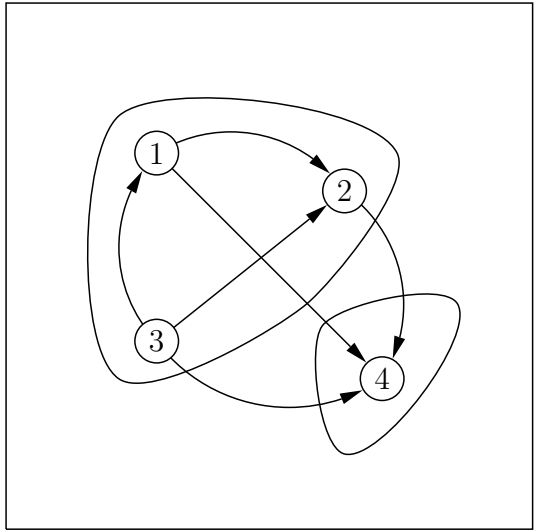

\begin{center}
\begin{graph}(7,7)
  \graphnodesize{.6}  

  \graphnodecolour{1}
  \roundnode{1}(2,5)
  \roundnode{4}(5,2)
  \roundnode{2}(4.5,4.5)
  \roundnode{3}(2,2.5)

  \freetext(2,5){1}
  \freetext(5,2){4}
  \freetext(4.5,4.5){2}
  \freetext(2,2.5){3}

  \dirbow{1}{2}{.2}
  \dirbow{2}{4}{.2}
  \dirbow{3}{1}{.2}
  \dirbow{3}{4}{-.2}

  \diredge{3}{2}
  \diredge{1}{4}

  \graphfillcolour{.8}
  \bubble{.2}{(1.5,5.5)(5.2,5)(4,3)(1.5,2)}
  \bubble{.2}{(4.2,2.7)(6,3)(4.5,1)}

  \filledareasfalse
  \area(0,0){(0,7)(7,7)(7,0)(0,0)}
\end{graph}
\end{center}
\caption{Coarse grained view of the thesaurus as a directed graph.
The region composed by subgraphs 1 to 3 is the \emph{source} and
subgraph 4 is referred as the \emph{sink}. The source contains:
 the \emph{normal source}, named as subgraph 1;
\emph{bridge source}, named as subgraph 2 and
\emph{absolute source}, here called subgraph 3.
}
\label{fig:g0}
\end{figure}

The directed graph concepts permit us to classify sets of terms according to its links properties as follows:
\begin{description}
\item[sink]  composed of the 73,046 terms with $k_{out} = 0$. 
For example: glucose, password, all-around, grape juice, send word, put to, lap dog, afterbirth; 

\item[source] are the 30,260 terms with at least one outgoing 
     link ($k_{out} > 0$), usually called main entries, entries, 
     head-words or root words. 
     The source can be divided into three categories;
\begin{description}
  \item[absolute source] is related to 877 terms without incoming links $k_{in} = 0$. 
       For example: rackets, grammatical, double quick, half moon, blinded;
  \item[normal source] are 29,333 terms that receive links and send links to other source and sink terms ($k_{out} > 0$ and $k_{in} > 0$). 
       For example: ablation, analogy, call out, factitious, laid low, make a deal; 
  \item[bridge source] they are the 16 terms without outgoing 
       links to source terms ($k_{out}(\mbox{source}) = 0$), listed: androgyny, Christian sectarians, 
       Congress, detector, electric meter, enzyme, Esperanto, et cetera, 
       Geiger counter, ghetto dwellers, harp, in fun, lobotomy, penicillin, 
       perversely, Senate;
\end{description}
\end{description}
These definitions are illustrated by the subgraphs 1--4 in 
Figure~\ref{fig:g0}.

\begin{figure}[htb]
\begin{center}
\includegraphics[width=.7\textwidth,angle=-90]{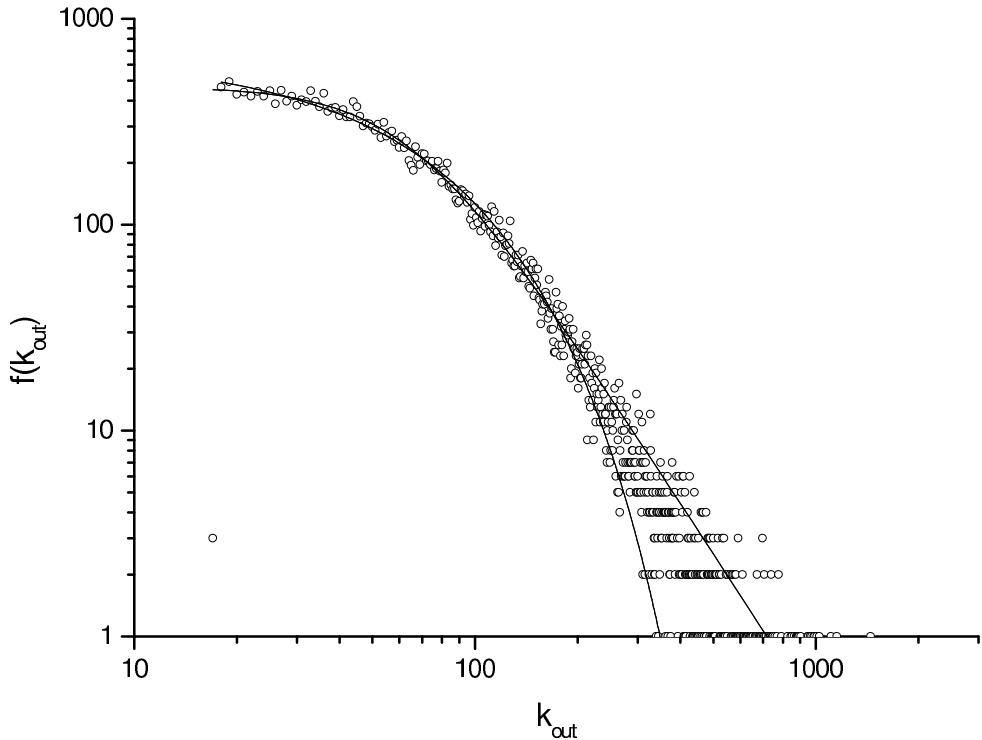}
\end{center}
\caption{The frequency of outgoing links $k_{out}$ (root words) 
is well described by Eq.~\ref{eq:kout} which is the rightwards curve,
 in contrast with curve of Eq.~\ref{eq:kout2}.
The point [$k_{out} = 17,f(k_{out}) = 3$] has been excluded in both 
fitting procedures. These words are: for good, for keeps, and grin.
}
\label{fig:g1}
\end{figure}

It is interesting to observe that the outgoing link distribution could be fitted by the scientific-papers citation frequency curve~\cite{albuquerque:2000}:   
\begin{equation}
\label{eq:kout2}
  f(k_{out}) = \frac{N_0}{\left[1 + (q-1) \lambda k_{out}\right]^{q/(q-1)}} \; ,
\end{equation}
with: $N_0 = 654 \pm 7$, $\lambda = (1.66 \pm 0.01) \times 10^{-2}$ and $q = 0.95 \pm 0.01$ ($r^2 = 0.993$ and $\chi^2 = 77$), see Fig.~\ref{fig:g1}.  
However, its limiting behavior for small $k_{out}$ is exponential instead of the measured stretched exponential. 
So, we propose another fitting function with power law behavior for
large $k_{out}$ and stretched exponential behavior for small
$k_{out}$\footnote{Note that introducing a new parameter one can write
  $f(x) = N_0/[1 + \lambda x^{\kappa}]^{q/(q-1)}$ which is the Burr XII distribution  function that appears as a result of a $q$-logarithm entropy maximization~\cite{brouers:2003} and generalizes both functions.}  
\begin{equation}
\label{eq:kout}
f(k_{out}) = \frac{N_0}{1 + \lambda k_{out}^{\kappa}}  \; ,
\end{equation}
with: $N_0 = 468 \pm 3$, $\lambda = (2.0 \pm 0.4) \times 10^{-5}$ and $\kappa = 2.55 \pm 0.03$ ($r^2 = 0.990$ and $\chi^2 = 99$), see Fig.~\ref{fig:g1}. 

For $\lambda k_{out}^{\kappa} \ll 1$ this curve  presents the same
behaviour of a stretched exponential~\cite{laherrere:1998}:
$f(k_{out}) = N_0 \exp (- \; k_{out}/\bar{k}_{out})^{\kappa}$, which
permits us to estimate the  mean value of outgoing links equal to
$\bar{k}_{out} = \lambda^{-1/\kappa} = 70 \pm 6$, which must be
compared to $83$ which is obtained from the thesaurus statistics. 
For $\lambda k_{out}^{\kappa} \gg 1$, the distribution of $k_{out}$ is a power law: $f(k_{out}) = (N_0/\lambda) k_{out}^{-\kappa}$.

\begin{figure}
\begin{center}
\includegraphics[width=.7\textwidth,angle=-90]{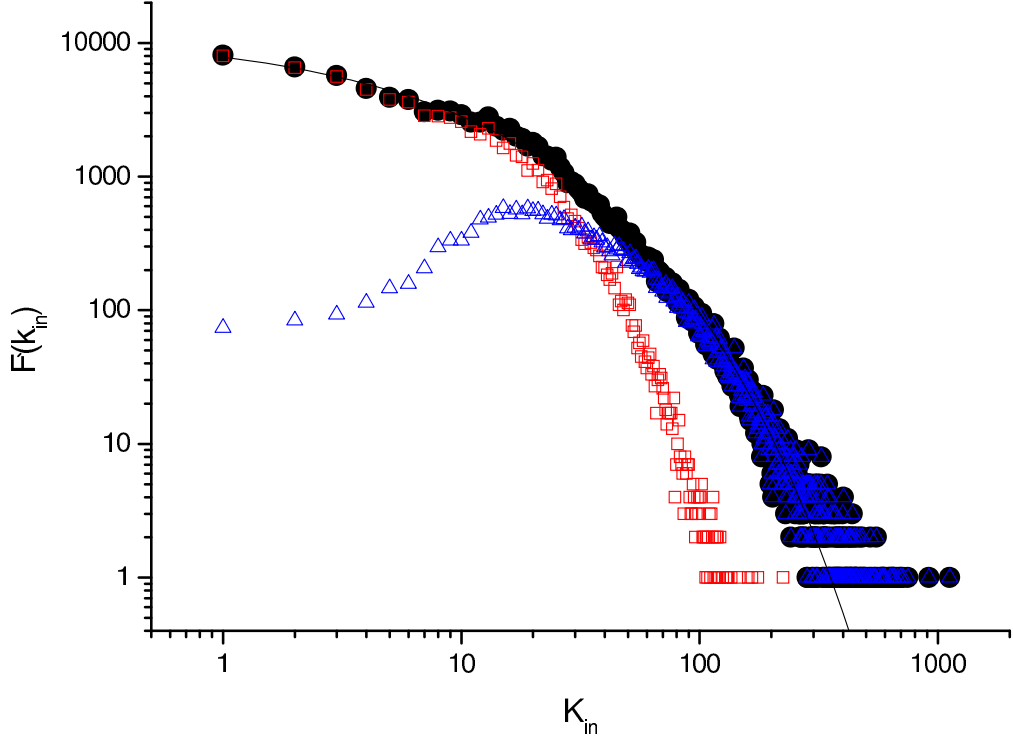}
\end{center}
\caption{Frequency of incoming links $k_{in}$ for all words ($\bullet$), root words ($\triangle$) and non-root words ($\Box$). 
The curve for  all words ($\bullet$) is well described by a stretched
exponential (line) expressed by Equation~\ref{eq:kin} ($N_0 = 12000
\pm 300$, $\bar{k}_{in} = 4.9 \pm 0.3$ and $\kappa = 0.52 \pm 0.01$)
which is dominated by non-root words for low $k_{in}$ values $(k_{in}
\le 10)$ and by root words for high $k_{in}$ values $(k_{in} \ge 100)$.}
\label{fig:g2}
\end{figure}

On the other hand, we show in Figure~\ref{fig:g2} that the frequency of words with a given number of incoming links ($k_{in}$) is very well described by the stretched exponential curve: 
\begin{equation}
\label{eq:kin}
f(k_{in}) = N_0 \exp \left(- \; \frac{k_{in}}{\bar{k}_{in}}\right)^{\kappa} \, , 
\end{equation}
where we have found $N_0 = 12000 \pm 300$, $\bar{k}_{in} = 4.9 \pm 0.3$
and $\kappa = 0.52 \pm 0.01$, ($r^2 = 0.993$ and $\chi^2 = 4.58$).  
We shall stress that a fitting curve of the type of Eq.~\ref{eq:kout} also describe this data if $\lambda$ is taken small enough.  
A simple approximation may be used as: $f(k_{in}) \propto \exp(-\sqrt{k_{in}})$. 
The low values of incoming links ($k_{in} < 10$) are dominated by non-root words while high values  ($k_{in} > 100$) are dominated by root words, as seen in Figure~\ref{fig:g2}.

\begin{figure}
\begin{center}
\includegraphics[width=.7\textwidth]{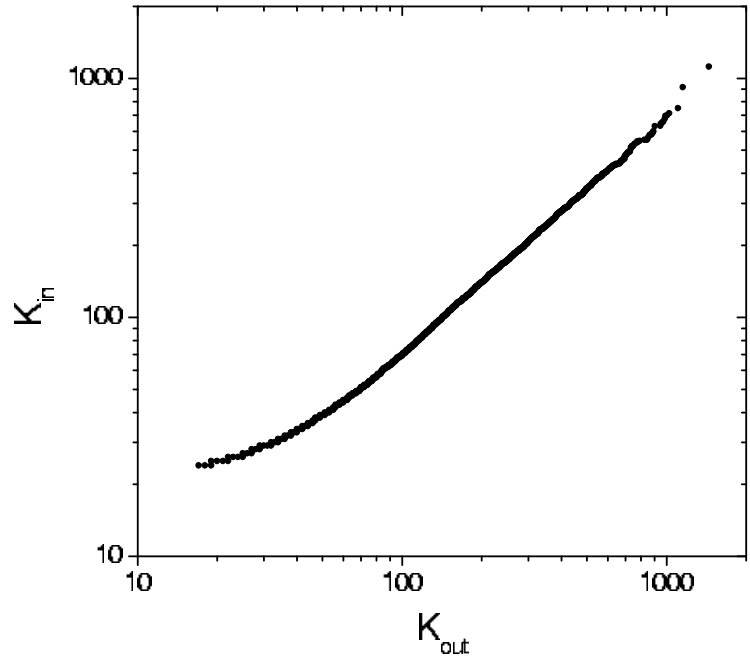}
\end{center}
\caption{The number of links $k_{in}$ and $k_{out}$ are ranked by decreasing values and plotted jointly to show the correlation.}
\label{fig:g3}
\end{figure}

Although empirically $f(k_{in})$ and $f(k_{out})$ are apparently
different, this may be due to a finite size database effect. 
This is suggested by a $k_{in} \times k_{out}$ plot (Fig.~\ref{fig:g3}) where $k_{in}$ and $k_{out}$ are ranked by decreasing values and plotted jointly to show the correlation between them.
From Fig.~\ref{fig:g3}, it is clear that a linear correlation occurs for $k > 100$. 
A perfect thesaurus should have a symmetric property $k_{in} = k_{out}$. 

As suggested by the above analysis, let Eq.~\ref{eq:kout} represent both the  distribution of outgoing and incoming links. 
If one takes the variable to be continuous, it is not hard to notice that Eq.~\ref{eq:kout} is the solution of the Ricatti's differential equation\footnote{Ricatti's differential equation is a particular instance of Bernoulli's differential equation~\cite{boyce_diprima}.}
\begin{equation}
y'(x) = - \; \frac{\kappa \lambda x^{\kappa-1}}{N_0} y^2(x) \; .
\end{equation}
This equation is known to represent contact processes such as the propagation of diseases~\cite{boyce_diprima}.  
Presently we are searching for a microscopic network growth model that has the Ricatti's equation as a continuum limit.

A thesaurus is a attempt to synthesize terms and their relationships  as natural as possible.
Nevertheless this trial is artificial and subjective.
Our work of treating the thesaurus as a directed graph has provided new insights into its macro structure. 
From this graph theoretical approach the counting of $k_{in}$ and $k_{out}$ could lead to a novel proposition of term arrangement and term connectivities in it.

The standard thesaurus classification is made according to word frequency in a corpus. 
But our approach suggested to rank the term related to a given root-word by its $k_{in}$ ranking or any other connectivity index.
For a finite thesaurus where the number of root-words is smaller than the total number of terms, we suggest that it is always possible to construct a new closed thesaurus by inversion between the initial $k_{out}>0$ root-words and $k_{in} > 0$ non-root words. 
The information in both cases are the same, but the latter leads to practical facilities, for instance, the fact that all words present in the thesaurus become root-words (the thesaurus becomes closed). 
We are submitting our closed version of Moby Thesaurus as a freeware database to the Gutenberg project~\cite{mthesaurus}. 

The authors are deeply grateful to Vera L\'ucia Coelho Villar 
from the Instituto Ant\^onio Houaiss de Lexicografia, Brazil, for the fruitful discussions
The authors thank stimulation discussion with F. Brouers , M. G. V. Nunes,
B. C. D. da Silva and
C. Tsallis.
This work has been partially funded by the Brazilian agencies: FAPESP, CAPES and CNPq. 



\end{document}